%% file: HIG-17-009_temp.tex
\pdfoutput=1

\documentclass[11pt,twoside,a4paper,cmspaper,final,collab]{cms-tdr}
\def\svnVersion{473120P}\def\cmsCernNoTag{CERN-EP-2018-127}\def\cmsCernDate{\today}\def\cmsMessage{Published in the Journal of High Energy Physics as \href{http://dx.doi.org/10.1007/JHEP08(2018)152}{\doi{10.1007/JHEP08(2018)152.}}}

\begin{document}\cmsNoteHeader{HIG-17-009}

\hyphenation{had-ron-i-za-tion}
\hyphenation{cal-or-i-me-ter}
\hyphenation{de-vices}
\RCS$HeadURL: svn+ssh://svn.cern.ch/reps/tdr2/papers/HIG-17-009/trunk/HIG-17-009.tex $
\RCS$Id: HIG-17-009.tex 471266 2018-08-07 17:35:09Z cvernier $

\cmsNoteHeader{HIG-17-009}
\title{Search for resonant pair production of Higgs bosons decaying to bottom quark-antiquark pairs in proton-proton collisions at 13\TeV}

\newcommand{\mH}{\ensuremath{m_\PH}\xspace}
\newcommand{\HbbHbb}{\ensuremath{\PH(\bbbar)\PH(\bbbar)}\xspace}
\newcommand{\XHbbHbb}{\ensuremath{\mathrm{X}\to\PH(\bbbar)\PH(\bbbar)}\xspace}
\newcommand{\ppXHbbHbb}{\ensuremath{\Pp\Pp\to\mathrm{X}\to\PH(\bbbar)\PH(\bbbar)}\xspace}

\date{\today}

\abstract{
A search for a narrow-width resonance decaying into two Higgs bosons, each decaying into a bottom quark-antiquark pair, is presented. The search is performed using proton-proton collision data corresponding to an integrated luminosity of 35.9\fbinv at $\sqrt{s}=13\TeV$ recorded by the CMS detector at the LHC. No evidence for such a signal is observed. Upper limits are set on the product of the production cross section for the resonance and the branching fraction for the selected decay mode in the resonance mass range from 260 to 1200\GeV.
}

\hypersetup{%
pdfauthor={CMS Collaboration},
pdftitle={Search for resonant pair production of Higgs bosons decaying to bottom quark-antiquark pairs in proton-proton collisions at 13 TeV},%
pdfsubject={CMS},%
pdfkeywords={CMS, physics, Higgs, bottom, di-Higgs}}

\maketitle

\section{Introduction}

The discovery of a Higgs boson ($\PH$)~\cite{Aad20121,Chatrchyan201230,Chatrchyan:2013lba}, with mass of 125\GeV~\cite{Aad:2015zhl,Sirunyan:2017exp} and properties consistent with the standard model (SM) of particle physics at the CERN LHC, motivates searches for resonances via their decays into Higgs bosons. Several theories for physics beyond the SM posit narrow-width resonances decaying into pairs of Higgs bosons ($\PH\PH$).
For instance, models with a warped extra dimension~\cite{Randall:1999ee} predict the existence of new particles such as the spin-0 radion~\cite{Goldberger:1999uk,DeWolfe:1999cp,Csaki:1999mp} and the spin-2 first Kaluza--Klein (KK) excitation of the graviton~\cite{Davoudiasl:1999jd,Csaki:2000zn, Agashe:2007zd}, which could decay to $\PH\PH$. These models have an extra warped spatial dimension compactified between two branes, with an exponential metric ${\kappa l}$, $\kappa$ being the curvature and $l$ the coordinate of the extra spatial dimension~\cite{Giudice:2000av}. The benchmark of the model is the ultraviolet cutoff of the theory, $\Lambda \equiv \sqrt{8\pi} e^{-\kappa l} {\Mpl}$, \Mpl being the Planck scale. In proton-proton ($\Pp\Pp$) collisions at the LHC, the graviton and the radion are produced primarily through gluon-gluon fusion and are predicted to decay to $\PH\PH$ with branching fractions of approximately 10 and 23\%, respectively~\cite{Oliveira:2014kla}.

Previous searches for resonant $\PH\PH$ production have been performed by the ATLAS and CMS Collaborations with $\Pp\Pp$ collisions at $\sqrt{s}=8$ and 13\TeV. The decay channels studied include $\bbbar\bbbar$~\cite{Khachatryan:2015yea,Aad:2015uka,Aaboud:2016xco}, $\bbbar\tau\tau$~\cite{Sirunyan:2017djm}, $\bbbar\gamma\gamma$~\cite{Khachatryan:2016sey,PhysRevLett.114.081802}, $\gamma\gamma \PW\PW$~\cite{PhysRevD.92.092004}, and $\bbbar \PW\PW$~\cite{Sirunyan:2017guj}.

This paper reports the results of a search for narrow-width resonances in the 260--1200\GeV mass range, decaying into a pair of Higgs bosons, each decaying into a pair of bottom quarks. The search is performed using $\Pp\Pp$ collision data collected at $\sqrt{s} = 13 \TeV$ with the CMS detector at the CERN LHC, corresponding to an integrated luminosity of 35.9\fbinv. The main challenge of this search is to discriminate the final-state signature of four bottom quark jets from the overwhelming multijet quantum chromodynamics (QCD) background. This is addressed by dedicated online selection criteria that include \cPqb\ jet identification and by a model of the multijet background that is tested in control regions of data. The analysis closely follows the approach adopted for the 8\TeV data~\cite{Khachatryan:2015yea} but the sensitivity for high resonance mass values is enhanced because of the significant increase in production cross section at 13\TeV, a new trigger strategy and a more efficient algorithm for identifying jets originating from bottom quarks.

\section{Detector and simulated samples}

The central feature of the CMS apparatus is a superconducting solenoid of 6\unit{m} internal diameter, providing a magnetic field of 3.8\unit{T}. A silicon pixel and strip tracker, a lead tungstate crystal electromagnetic calorimeter, and a brass and scintillator hadron calorimeter, each composed of a barrel and two endcap sections, reside within the solenoid. Forward
calorimeters extend the pseudorapidity ($\eta$)~\cite{Chatrchyan:2008zzk}
coverage provided by the barrel and endcap detectors. Muons are detected in gas-ionization chambers embedded in the steel flux-return yoke outside the solenoid. A more detailed description of the CMS detector, together with a definition of the coordinate system used and the relevant kinematic variables, can be found in Ref.~\cite{Chatrchyan:2008zzk}

Simulated samples of signal events are produced using various Monte Carlo (MC) event generators, with the CMS detector response modeled using the \GEANTfour~\cite{GEANT4} program. To model the production of a generic narrow-width spin-0 resonance, we use an MC simulation of the bulk radion produced through gluon fusion. The momenta and angular distributions for decay products of a spin-2 resonance are distinct from those for a spin-0 resonance, and result in different kinematic distributions. Therefore, we evaluate the signal efficiencies for a narrow-width spin-2 resonance from a separate simulation of the first excitation of a bulk KK graviton produced through gluon fusion and forced to decay to a pair of Higgs bosons with the parameters reported in Ref.~\cite{RadionMC}. Bulk graviton and radion signal events are simulated with masses in the range 260--1200\GeV and widths of 1\MeV (narrow-width approximation), using the {\MGvATNLO 2.3.3}~\cite{MG5_aMCNLO} event generator at leading order (LO). The resonance is forced to decay into a pair of Higgs bosons which in turn decay into \bbbar. The parton distribution function (PDF) set NNPDF3.0~\cite{Ball:2014uwa} with LO accuracy is used.
The showering and hadronization of partons are simulated with \PYTHIA~8.212~\cite{Pythia8p2}.

During the 2016 data-taking period the average number of $\Pp\Pp$ interactions per bunch crossing was approximately 23. The simulated samples include these additional $\Pp\Pp$ interactions, referred to as pileup interactions (or pileup), that overlap with the event of interest in the same bunch crossing. Simulated events are weighted to match the number of $\Pp\Pp$ interactions per event in data.

\section{Event reconstruction}
\label{sec:Reconstruction}

The particle-flow (PF) algorithm~\cite{Sirunyan:2017ulk} is used to reconstruct and identify individual particle in an event with
an optimized combination of information from
the various elements of the CMS detector. The algorithm identifies each reconstructed particle (PF candidate) as an electron, a muon, a photon, or a charged or neutral hadron.

The reconstructed vertex with the largest value of summed physics-object transverse momentum squared ($\pt^2$) is taken to be the primary $\Pp\Pp$ interaction vertex. The physics objects are the jets, clustered using the jet finding algorithm~\cite{antikt,Cacciari:fastjet1} with the tracks assigned to the vertex as inputs, and the associated missing transverse momentum, taken as the negative vector sum of the \pt of those jets. This vertex is used for all the objects in the event reconstructed with the PF algorithm.

Jets are reconstructed from PF candidates using the anti-\kt clustering algorithm~\cite{antikt}, with a distance parameter of 0.4, as implemented in the \FASTJET package~\cite{Cacciari:fastjet1,Cacciari:fastjet2}. Jet identification criteria are also applied to reject jets originating from detector noise. The average neutral energy density from pileup interactions is evaluated from PF objects and subtracted from the reconstructed jets~\cite{Cacciari:2007fd}. Jet energy corrections are derived from the simulation, and are confirmed with in situ measurements of the energy balance in dijet and photon+jet events~\cite{Khachatryan:2016kdb}.

Jets are identified as originating from $\cPqb$ quarks (``$\cPqb$ jets") using the \textsc{DeepCSV}~\cite{DeepCSV} discriminator, a
new \cPqb\ tagging algorithm based on a deep neural network with four hidden layers~\cite{Guest:2016iqz}. The
\textsc{DeepCSV} discriminator employs the same set of observables as those used by the combined secondary
vertex (CSV) algorithm~\cite{Chatrchyan:2012jua,DeepCSV}, except that the track selection is expanded to
include up to six tracks, further improving the \cPqb\ jet identification. The operating point chosen corresponds to a 1 (12)\% rate for misidentifying a light-flavor ({\cPqc}-flavor) jet as a \cPqb\ jet. The \cPqb\ tagging efficiency for jets with $\pt$ in the 30--150\GeV range is
approximately 69\% and gradually decreases for lower and higher jet \pt~\cite{DeepCSV}.

\section{Event selection}
\label{sec:EventSelection}

The search for a narrow-width \XHbbHbb ~resonance is performed for mass values in the 260--1200\GeV range. The angular distributions for the decay products of such a resonance vary substantially over this range. In order to increase the sensitivity of this search, different criteria are used for events in two distinct mass regions: the low-mass region (LMR), for resonance masses from 260 to 620\GeV, and the medium-mass region (MMR), for masses from 550 to 1200\GeV. The boundary between the LMR and the MMR is at 580\GeV. It has been chosen by optimizing for the expected sensitivity and
takes into account the uncertainties associated with the background modeling. The mass range above 1200\GeV (high-mass region) is not covered by this search. Above 900\GeV, the Higgs bosons have a momentum considerably higher than their mass and the Higgs to \bbbar decays are reconstructed more efficiently as single hadronic jets with a larger anti-\kt distance parameter (0.8)~\cite{B2G-16-026}.

Events are selected online by combining two different trigger selections to identify \cPqb\ jets, both using the CSV algorithm. For the first trigger selection, four jets with $\pt >30\GeV$ and $\abs{\eta} < 2.4$ are required. The latter requirement ensures that the jet lies within the tracker acceptance. Of those four jets, two are required to have $\pt >90\GeV$ and at least three jets are required to be tagged as \cPqb\ jets. The second trigger selection requires four jets with $\pt > 45\GeV$ and at least three of those jets identified as \cPqb\ jets.

Events are selected offline by requiring at least four \cPqb\ tagged jets with $\pt >30\GeV$ and $\abs{\eta} < 2.4$. The selected jets are combined randomly into pairs to form two Higgs boson candidates with masses $m_{\PH_1}$ and $m_{\PH_2}$.
For the LMR, $\PH\PH$ candidates are chosen from the four selected jets such that $\abs{m_{\PH} - 120 \GeV} < 40 \GeV$ for each candidate Higgs boson.
For the MMR the \PH\ candidates are selected using $\Delta R = \sqrt{\smash[b]{(\Delta\eta)^2+(\Delta\phi)^2}}$ less than 1.5, where $\Delta \eta$ and $\Delta \phi$ are the differences in the pseudorapidities and azimuthal angles (in radians) of the two jets.

In the two-dimensional space defined by the reconstructed masses of the two Higgs boson candidates, ${\PH_1}$ and $\PH_2$, a circular signal region (SR) is defined with $\mathcal{R}<1$, where $\mathcal{R}$ is defined as:
\begin{equation}
\mathcal{R}  = \sqrt{\left(\frac{m_{\PH_1}-\mathcal{M}}{r}\right)^{2} + \left(\frac{m_{\PH_2}-\mathcal{M}}{r}\right)^{2}}
\label{eq:chi2}
\end{equation}
The central mass value ($\mathcal{M}$) is the average of the means of the $m_{\PH_1}$ and $m_{\PH_2}$ distributions for simulated signal events and the parameter $r$ is set to 20\GeV. The centers of these circular regions have been determined separately for the LMR and MMR and found to be 120 and 125\GeV, respectively.
If there are multiple $\PH\PH$ candidates in an event, the combination that minimizes $\mathcal{R}^2$ is used.

\begin{figure}[htb!]
\centering
\includegraphics[width=0.45\textwidth]{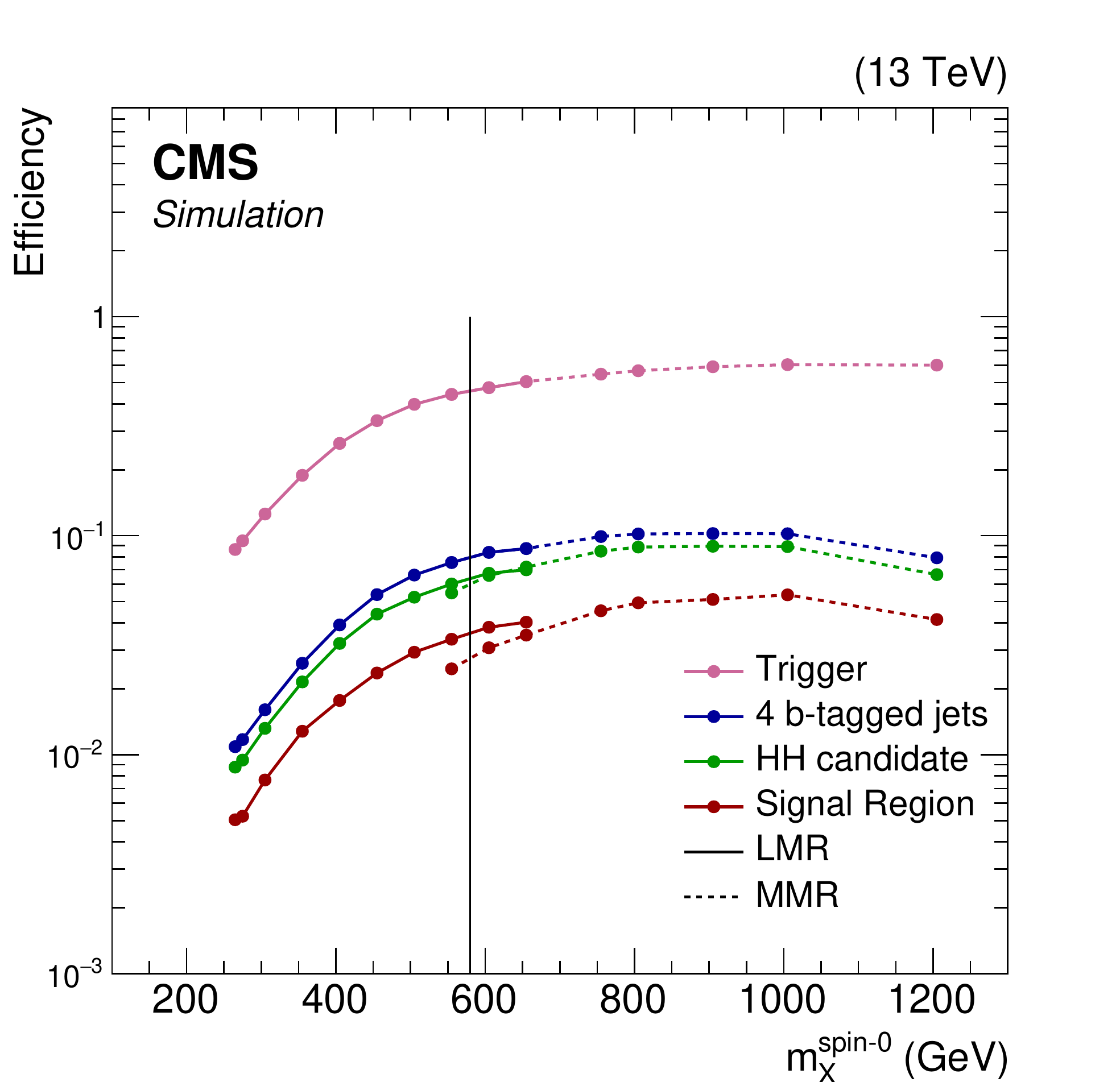}
\includegraphics[width=0.45\textwidth]{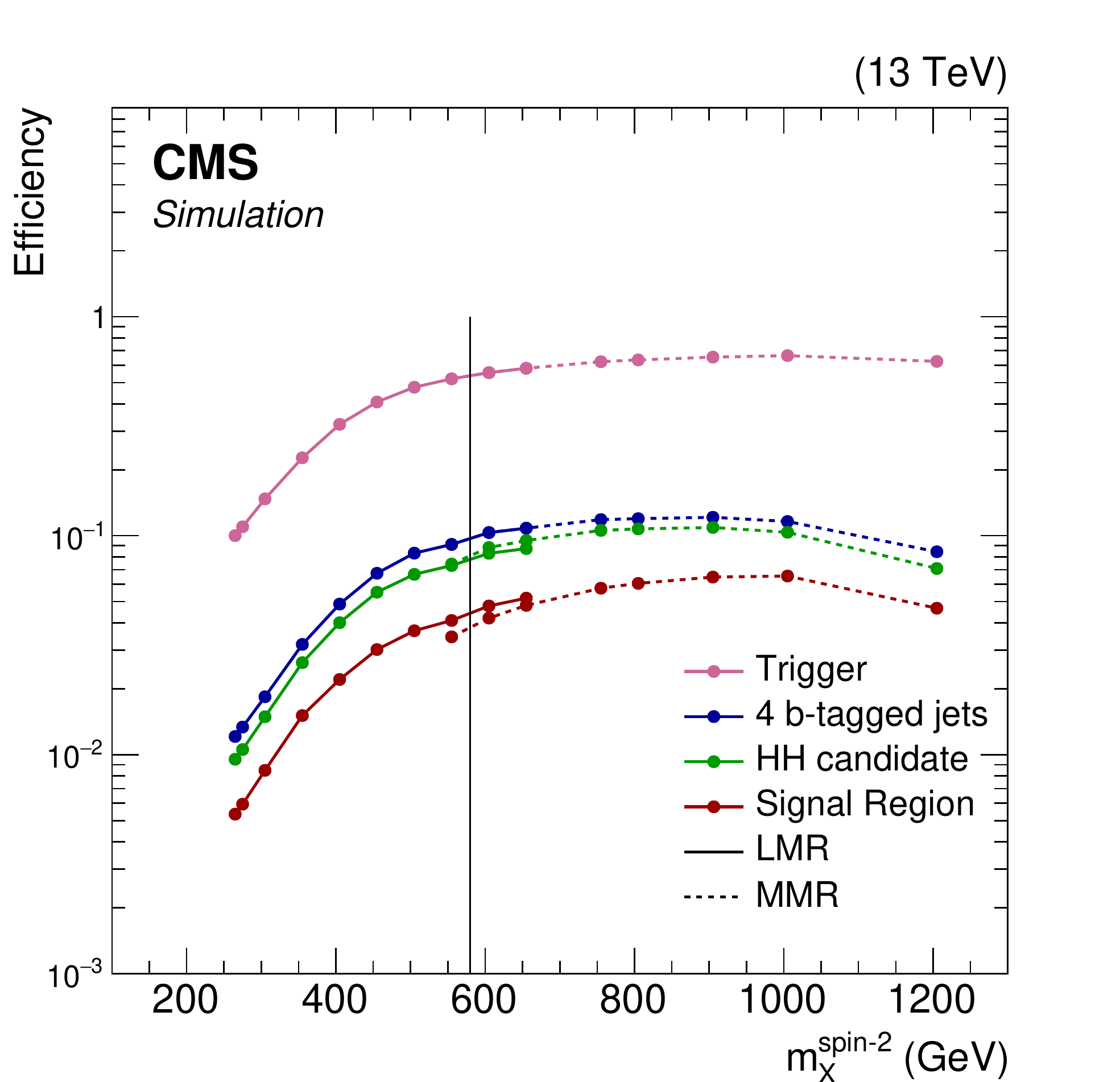}

\caption{The selection efficiency for simulated \XHbbHbb events for a spin-0 radion (left) and a spin-2 bulk KK-graviton (right), at different stages of the event selection for each mass hypothesis, for the low-mass region (solid) and the medium-mass region (dashed). The vertical line at 580\GeV corresponds to the transition between the LMR and MMR regions. }
\label{fig:CutFlow}
\end{figure}

After these event selection criteria are applied, the
dijet invariant mass resolution for $m_{\PH_1}$ and $m_{\PH_2}$ is
approximately 10--13\%, depending on the \pt\ of the reconstructed Higgs boson,
with a few percent shift in the value of the mass peak, relative to 125\GeV. The
Higgs boson mass resolution is further improved by applying multivariate
regression techniques similar to those used in the searches for SM Higgs bosons decaying to \bbbar in CMS~\cite{Chatrchyan:1610290,Khachatryan:2015bnx}. The regression estimates a correction
that is applied after the standard CMS jet energy
corrections~\cite{cmsJEC,Khachatryan:2016kdb}, and it is computed for individual \cPqb\ jets to improve
the accuracy of the measured energy with respect to the \cPqb\ quark energy.  To this
end, a specialized boosted decision tree~\cite{Hocker:2007ht} is trained on simulated $\cPqb$ jets from \ttbar events, with inputs that include observables related to the jet structure and \cPqb\ tagging information.
The average improvement in the Higgs boson mass resolution, measured with simulated signal samples, is 6--12\%, depending on the \pt\ of the
reconstructed Higgs boson.  The use of the regression technique increases the
sensitivity of the analysis by 5--20\% depending on the mass hypothesis.
The regression technique is validated with data samples of $\cPZ\to(\Pe\Pe,\mu\mu)$ events with two \cPqb\ tagged jets and in
$\ttbar$-enriched samples~\cite{Chatrchyan:1610290}.
The cumulative selection efficiencies of the selection criteria described above for the graviton and radion signal benchmarks are reported in Fig.~\ref{fig:CutFlow}.

\begin{figure}[htb!]
\centering
\includegraphics[width=0.75\textwidth]{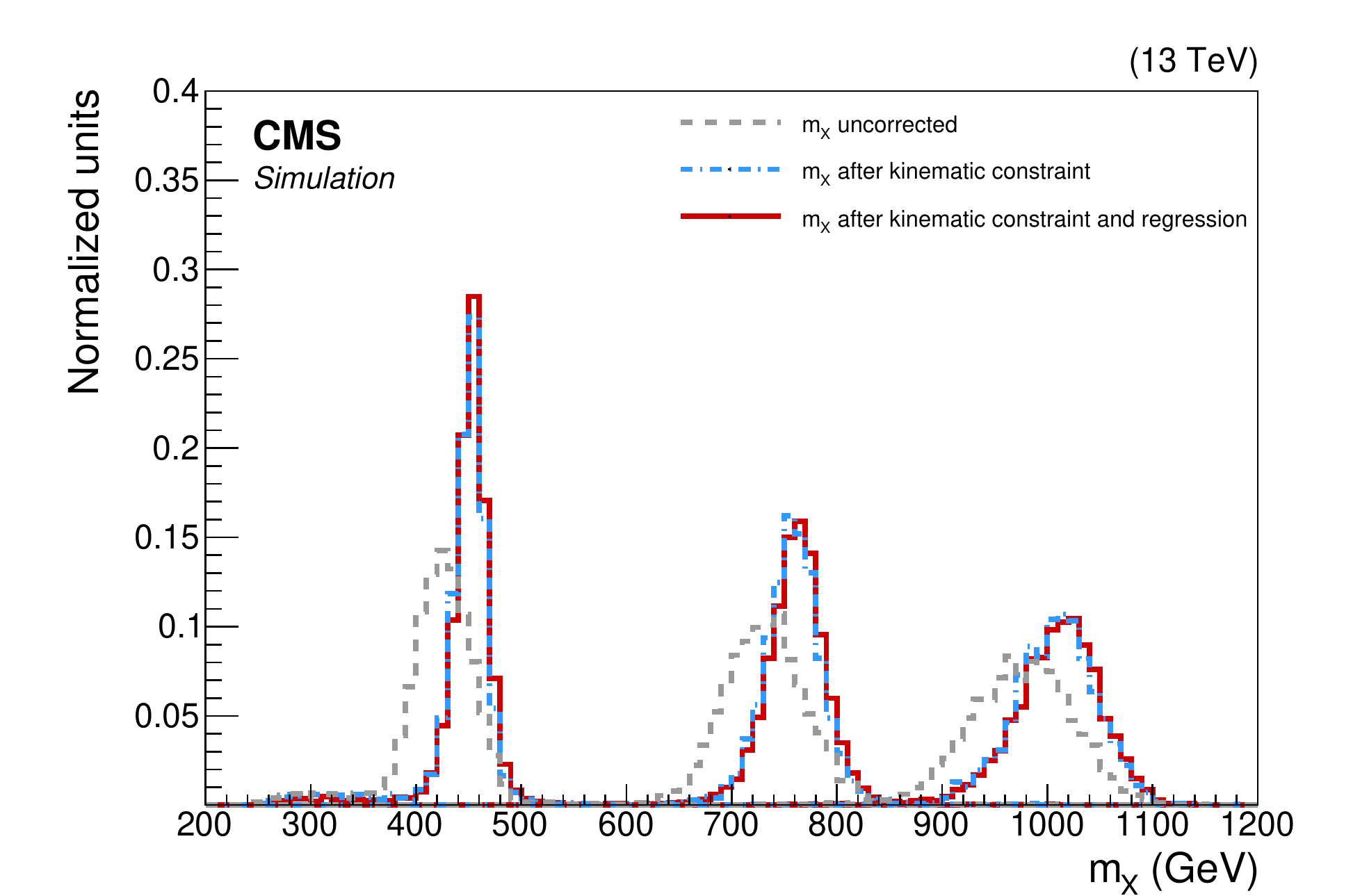}
\caption{The $m_{\mathrm{X}}$ distribution for simulated signal events (spin-2 bulk KK-graviton) after the event selection criteria for the 450, 750, and 1000\GeV mass hypotheses, with and without the correction obtained by constraining $m_{\PH}$ (kinematic constraint) and the specific \cPqb\ jet energy corrections (regression). The distributions are normalized so that the area under the curve for each mass is the same}
\label{fig:SignalShapes}
\end{figure}

The reconstructed resonance mass ($m_{\mathrm{X}}$), computed as the invariant mass of $\PH_1$ and $\PH_2$, is displayed for simulated signal events with different mass hypotheses in Fig.~\ref{fig:SignalShapes}.
In order to improve the resolution for the resonance mass, the momenta of the reconstructed $\cPqb$ quarks are corrected by constraining the invariant mass of the Higgs boson candidates to be 125\GeV. Since jet direction is reconstructed with better resolution than jet \pt, this constraint mainly benefits the latter. The improvement in resolution for the reconstructed signal resonance ranges from 20 to 40\% depending on the mass hypothesis, resulting in an improvement of the sensitivity by 10--20\%.
The shift in the reconstructed $m_{\mathrm{X}}$, due to these corrections is also shown in Fig.~\ref{fig:SignalShapes}. The mass shift is linear in $m_{\mathrm{X}}$ and is caused by the asymmetry of the corrections due to the jet momentum resolution across the \pt range considered.

\section{Signal and background modeling}
\label{sec:Background}

To search for signal events with various mass hypotheses we fit the $m_{\mathrm{X}}$ distribution for data events in the SR to the sum of two parametric models, one for the signal and one for the SM background, which is dominated by multijet production.
This procedure is performed independently in three regions: two within the LMR, as described below, and the MMR. Quantum interference between the signal and non-resonant SM background processes is neglected.

The parametric signal models are built by fitting the $m_{\mathrm{X}}$ distributions obtained from simulated signal events. The shape of the signal $m_{\mathrm{X}}$ distribution is different for the LMR and the MMR, as a result of the different event selection criteria. In the LMR, a sum of two Gaussian functions is used, to account for the effect of reconstructing the two Higgs boson candidates using jets not originating from their decays. This occurs in about 30\% of the events for the lowest resonance mass hypothesis, decreasing to about 5\% at the highest resonance mass hypothesis. Five parameters are required, the mean and width of the two Gaussian functions and the ratio of their integrals. In the MMR, the signal is modeled with a function that has a Gaussian core smoothly extended on both sides via exponential tails.
This requires two parameters for the mean and width of the Gaussian function, and two parameters for the exponential tails~\cite{Khachatryan:2015yea}.

\begin{figure}[tb]
\centering
\includegraphics[width=0.5\textwidth]{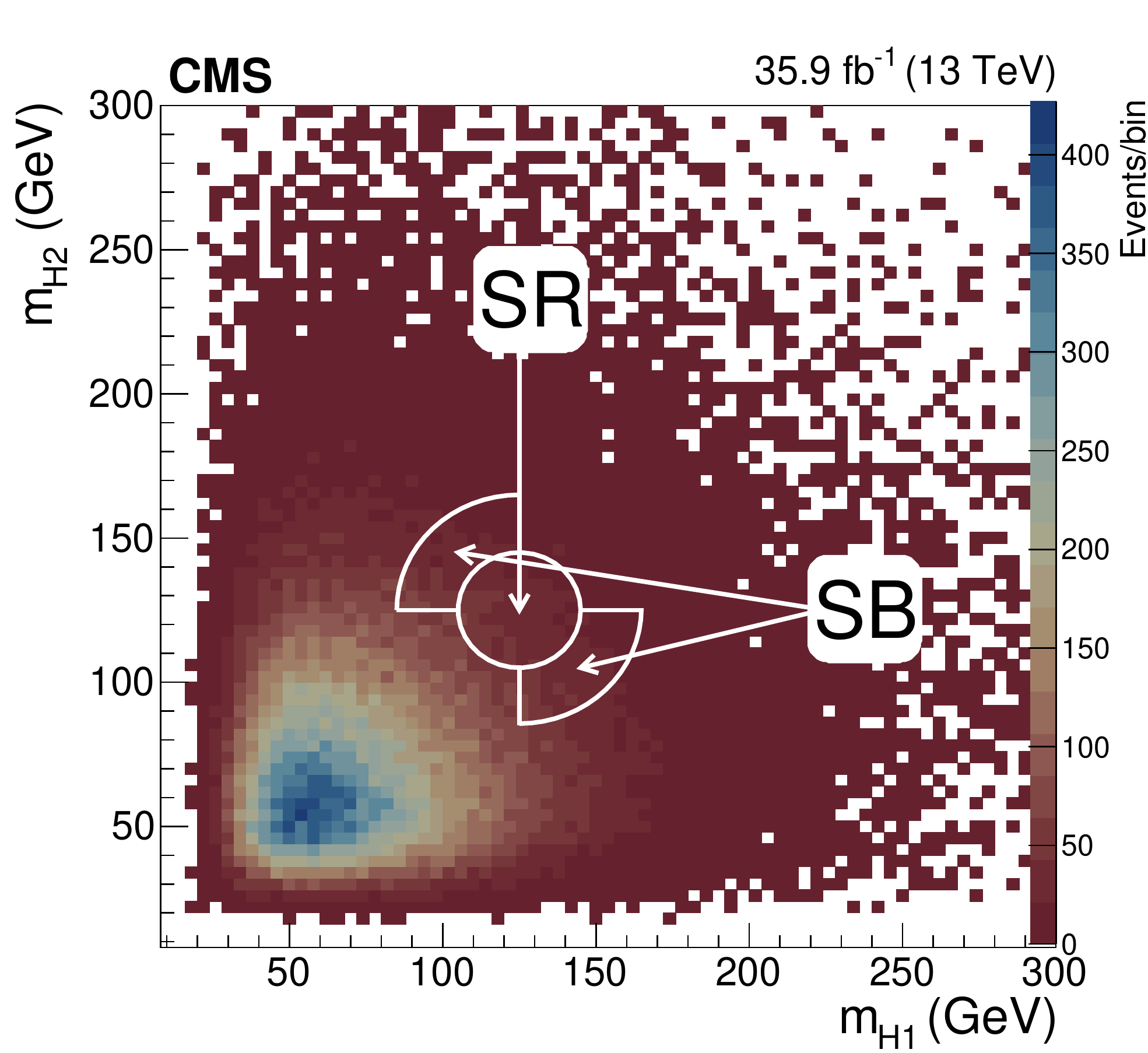}
\caption{Definition of the SR and the SB in the ($m_{\PH_1}$, $m_{\PH_2}$) plane used to motivate and validate the parametric model for the multijet background. The quantities $m_{\PH_1}$ and $m_{\PH_2}$ are the two reconstructed Higgs boson masses after applying the multivariate regression described in Section~\ref{sec:EventSelection}. Data corresponds to a selection in the MMR.}
\label{fig:Strategy}
\end{figure}

The parametric function used to model the background $m_{\mathrm{X}}$ distribution is obtained from control regions in data in which events are expected to have kinematic properties similar to events in the SR.
These control regions are: (i) sideband regions (SB) defined as $1 < \mathcal{R} < 2$ and $(m_{\PH_1} - \mathcal{M})(m_{\PH_2} - \mathcal{M})< 0$ in the $m_{\PH_1}-m_{\PH_2}$ plane (as shown in Fig.~\ref{fig:Strategy}), and (ii) events in the SR and both SB regions that do not qualify as signal events due to a requirement that one of the four jets used to reconstruct $m_{\mathrm{X}}$ is not identified as a \cPqb\ jet (anti-tag selection). For the LMR selection, the $m_{\mathrm{X}}$ distributions in all control regions exhibit a kinematic turn-on, due to the trigger requirements, followed by a smoothly falling distribution at the larger $m_{\mathrm{X}}$ values. The $m_{\mathrm{X}}$ distributions in all control regions for the MMR selection exhibit a similar smoothly falling shape with increasing $m_{\mathrm{X}}$. This feature allows the adoption of a common form of the parametric background model in the SR and control regions. The parameters of the background model are determined by the fit used to extract a possible signal. In the following we describe the derivation of the background model from the different control regions. For the LMR, it is difficult to model simultaneously the turn-on and the tail of the $m_{\mathrm{X}}$ distribution. Thus, the modeling of the background is split into two ranges, LMR1 and LMR2, with two different parametric models to accommodate either a turn-on shape or a falling distribution, respectively.

\begin{figure}[htb!]
\centering
\includegraphics[width=0.49\textwidth]{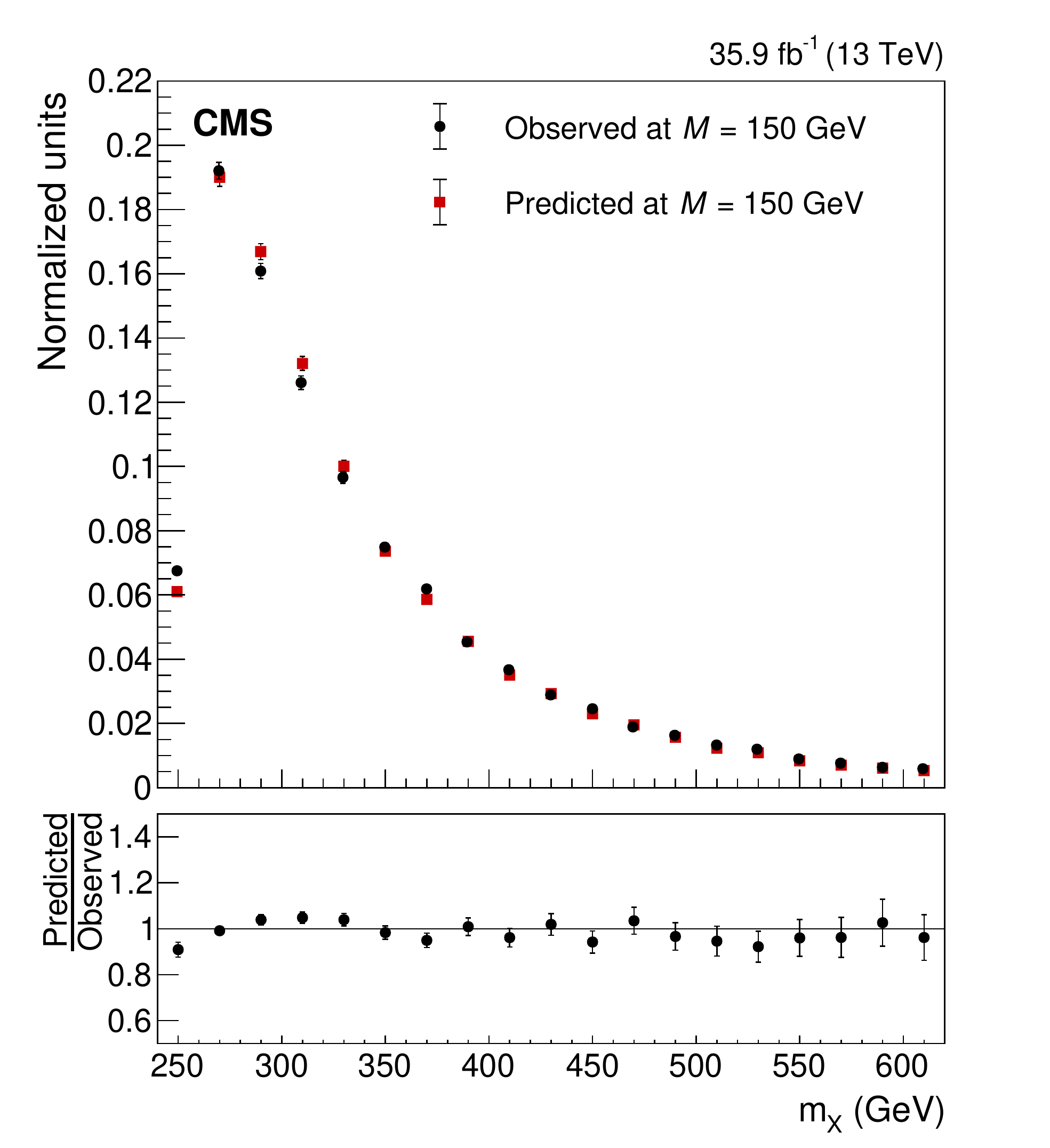}
\caption{The predicted $m_{\mathrm{X}}$ distribution in data for the LMR (squares) and the observed distribution in a SR (circles) centered at a Higgs boson mass of 150\GeV.}
\label{fig:MMMM}
\end{figure}

An indirect method of determining the boundary between LMR1 and LMR2 is used to avoid revealing the $m_{\mathrm{X}}$ distribution in the SR before fitting for a possible signal. The method uses two selections in the SR and SB, as described in Table~\ref{tab:ABCDmethod}. For the background, the shape of the $m_{\mathrm{X}}$ distribution in the SR (A) is predicted from the shape of the $m_{\mathrm{X}}$ distribution in the SR using the anti-tag selection (C) but reshaped by the ratio, as a function of $m_{\mathrm{X}}$, between the $m_{\mathrm{X}}$ distribution in the SB with the nominal selection (B) and the $m_{\mathrm{X}}$ distribution in the SB with the anti-tag selection (D), so that $\mathrm{A'=C {B}/{D}}$. To test the validity of this method, new SR and SB regions, centered at 150\GeV, are defined along the diagonal in Fig.~\ref{fig:Strategy} in a similar fashion to those centered at 120\GeV. In that case the prediction $\mathrm{E'=G {F}/{H}}$ for the $m_{\mathrm{X}}$ distribution in the SR region centered at 150\GeV can be tested by directly comparing to the $m_{\mathrm{X}}$ distribution in that region. Figure~\ref{fig:MMMM} shows the agreement between this prediction and the observed distribution, validating the use of this method for the SR distribution centered at 120\GeV. The boundary between LMR1 and LMR2 is then set at $m_{\mathrm{X}} =310$\GeV, by optimizing for the expected sensitivity.

\begin{figure}[tbp]
\centering
\includegraphics[width=0.49\linewidth]{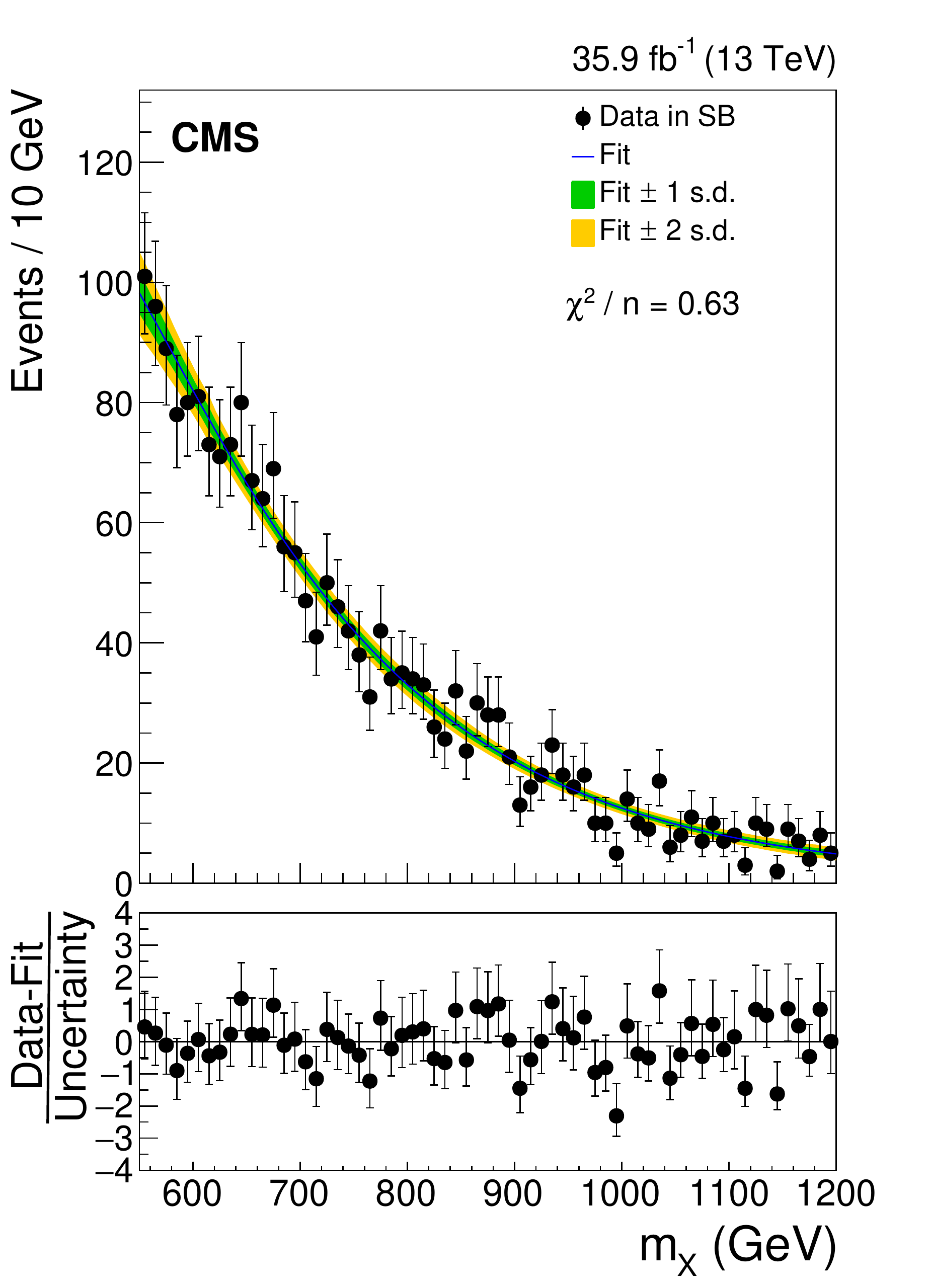}
\caption{The m$_X$ distribution in the SB of the MMR and the fit to the background multijet shape are shown. The shaded regions correspond to variations of $\pm$1 and $\pm$2 standard deviation (s.d.) in this parametrized form. Here $n$ is the number of degrees of freedom in each fit. The lower panel shows the difference between the data and the fits, divided by the uncertainty in the number of data events. }
\label{fig:MMMMbar}
\end{figure}

The signal and background are evaluated independently in LMR1$=\left[260,310\right]$\GeV, LMR2$=\left[310,580\right]$\GeV and the MMR$=\left[580,1200\right]$\GeV. For LMR1, a function with a Gaussian core smoothly extended to an exponential tail on the high side is fitted to the $m_{\mathrm{X}}$ distribution~\cite{Khachatryan:2015yea}, while the function defined in Ref.~\cite{Ikeda:1999aq} Eq.~(9) is used in LMR2 and the MMR. This function was originally used to describe a Compton spectrum and has three free parameters describing the mean, width and extent of the right side tail. In each case, the goodness of the fit, characterized by the $\chi^2$ per degree of freedom, is found to be reasonable. The fit of the background model on the SB for the MMR is shown in Fig~\ref{fig:MMMMbar}.

In the absence of a theoretical prediction for the background model, an alternative one based on the Crystal Ball function~\cite{Oreglia:1980cs} has been studied, and the difference between them drives the assessment of a systematic uncertainty due to the choice of the particular model. Pseudo-datasets are generated from the alternative function and fitted with the nominal functions to compute biases in the reconstructed signal strength. This procedure is performed for each mass hypothesis and the corresponding biases range between 30--80\unit{fb} for the LMR, and 0.1--4\unit{fb} for the MMR.

The SM top quark pair production in the SR is estimated from simulation to contribute up to 10 and 15\% of the selected events in the LMR and the MMR, respectively.  Since the $\ttbar$ contribution to the background exhibits a shape very similar to that for the multijet process, it is implicitly included in the data driven estimate.

\begin{table}[!htb]
\centering
\topcaption{Definitions of the control regions we used to test the functional form as described in the text.}
\begin{tabular}{lcccc}
        & \multicolumn{2}{c}{$\mathcal{M}={120}\GeV$} & \multicolumn{2}{c}{$\mathcal{M}=150\GeV$}\\
\hline
                        &       SR     &       SB     &       SR     &       SB     \\
{4 \cPqb\ tagged jets}            &       A       &       B       &       E       &       F       \\
{anti-tag}        &       C       &       D       &       G       &       H       \\
\end{tabular}
\label{tab:ABCDmethod}
\end{table}

\section{Systematic uncertainties}
\label{sec:Systematics}
Sources of systematic uncertainties that affect the signal yields are listed in Table~\ref{tab:Systematics}.
The signal yield for a given production cross section is affected by a 2.5\% systematic uncertainty in the measurement of the integrated luminosity at CMS~\cite{CMS-PAS-LUM-17-001}. The uncertainty in the signal normalization caused by the choice of the PDF set~\cite{Butterworth:2015oua} contributes up to 3.5\%. The jet energy scale~\cite{cmsJEC,Khachatryan:2016kdb} is varied within one standard deviation as a function of jet \pt and $\eta$, and the efficiency of the selection criteria recomputed. The signal efficiencies are found to vary by up to 2.9\%. The effect of the uncertainty in the jet energy resolution~\cite{cmsJEC,Khachatryan:2016kdb} is evaluated by smearing the jet energies according to the measured uncertainty. This causes variations in the signal efficiency of between 0.9 and 2.1\%. The trigger efficiency is evaluated in a $\ttbar$-enriched control sample in which an isolated muon is required in addition to the four \cPqb\ tagged jets with $\pt >30\GeV$ and $\abs{\eta} < 2.4$ demanded in the analysis. The efficiency of each online kinematic and \cPqb\ tagging requirement is evaluated separately and then the efficiency of all the trigger selection criteria is computed in terms of conditional probability. The associated systematic uncertainties are found to impact signal efficiencies at the level of 5--9\%.  Signal yields are corrected to match the \cPqb\ tagging efficiency measured in data~\cite{DeepCSV}. The associated uncertainty is evaluated to be about 6--8\%.

The systematic uncertainty associated with the choice of the parametric background model is evaluated as described in Section~\ref{sec:Background}. We account for the bias as a signal-shaped systematic uncertainty in the background model with normalization centered at zero and a Gaussian uncertainty with standard deviation equal to the bias. The impact on the expected limit ranges between 0.3--1.5\%.

\begin{table}[tbp]
\centering
\topcaption{Impact of systematic uncertainties on the signal efficiencies in the LMR and the MMR.}
\label{tab:Systematics}
\setlength{\tabcolsep}{0.90pc}
\begin{tabular}{lcccc}
Source of                & {Impact in LMR (\%)} & {Impact in MMR (\%)} \\
systematic uncertainty   & Signal         & Signal         \\
\hline
Luminosity		 & 2.5            & 2.5 \\
Jet energy scale         & 0.2--1.8     & 0.9--2.9         \\
Jet energy resolution     & 0.9--2.1          & 1.0--1.5         \\
\cPqb\ tagging scale factor   &  6.5--6.9              & 6.9--8.6             \\
Trigger efficiency     & 6.4--9.0         & 5.3--7.0        \\
PDF  &			 1.5--2.2     & 2.1--3.5 \\
\end{tabular}
\end{table}

\section{Results}
\label{sec:Results}

\begin{figure}[htbp!]
\centering
\includegraphics[width=0.49\linewidth]{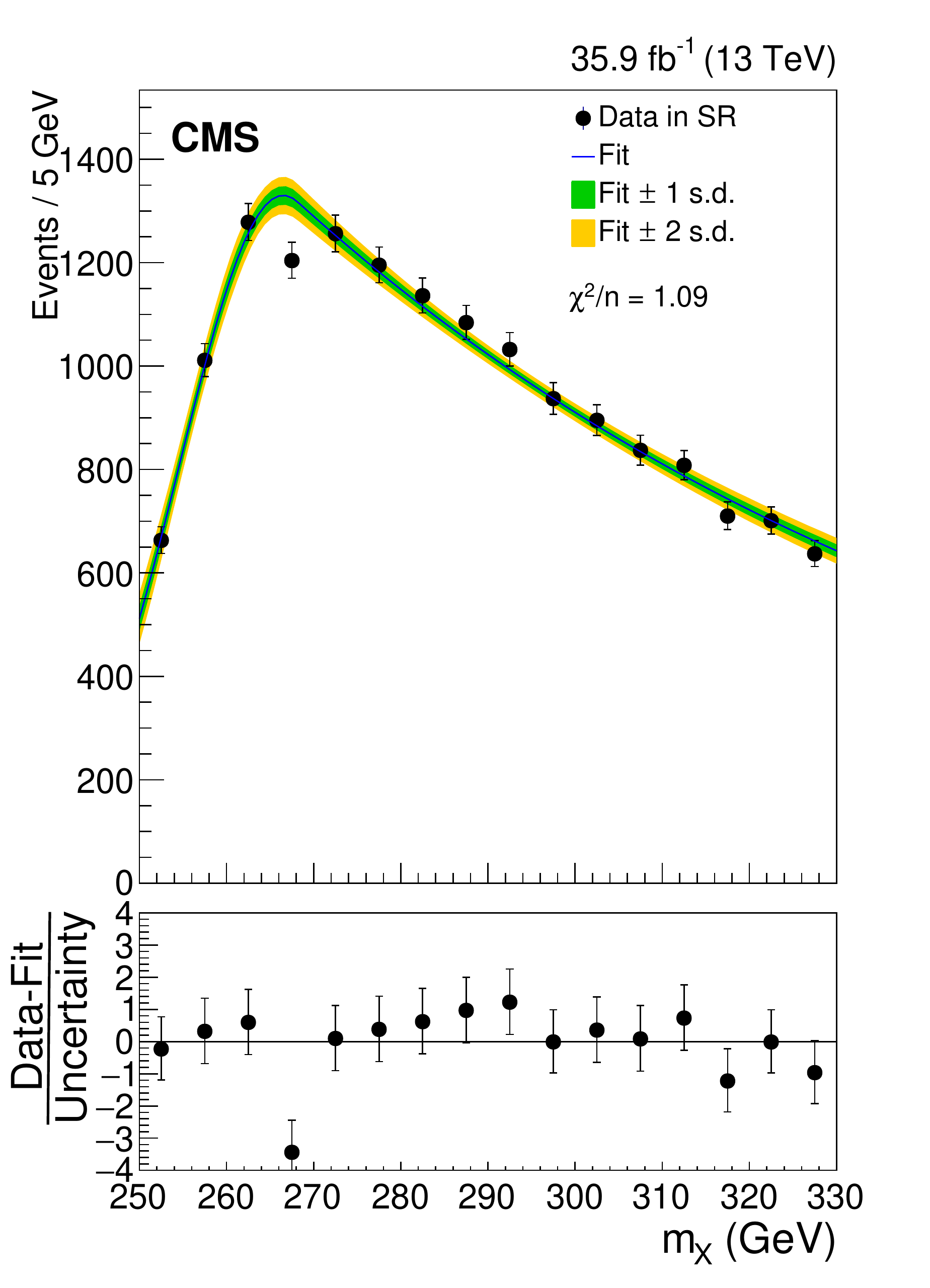}
\includegraphics[width=0.49\linewidth]{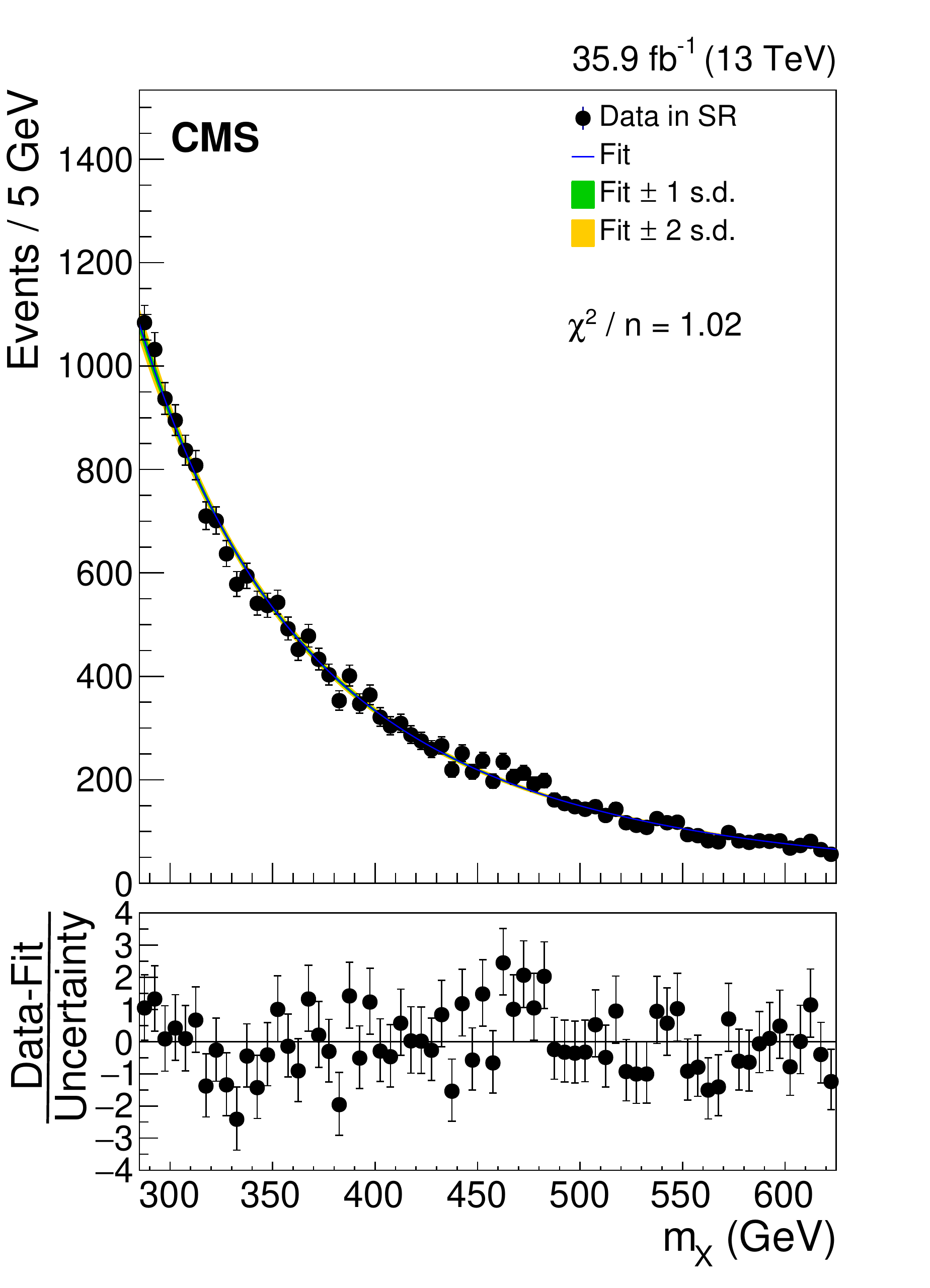}
\caption{The $m_{\mathrm{X}}$ distributions for LMR1 (left) and LMR2 (right) in the SR. These distributions are fitted in the two ranges to the reference model. A fit to the background-only hypothesis, multijets, is shown. The shaded regions correspond to variations of $\pm$1 and $\pm$2 standard deviation (s.d.) in the parametrized form.
Here $n$ is the number of degrees of freedom in each fit. The lower panels show the difference between the data and the fits, divided by the uncertainty in the number of data events. }
\label{fig:unblinded}
\end{figure}

\begin{figure}[htbp!]
\centering
\includegraphics[width=0.49\linewidth]{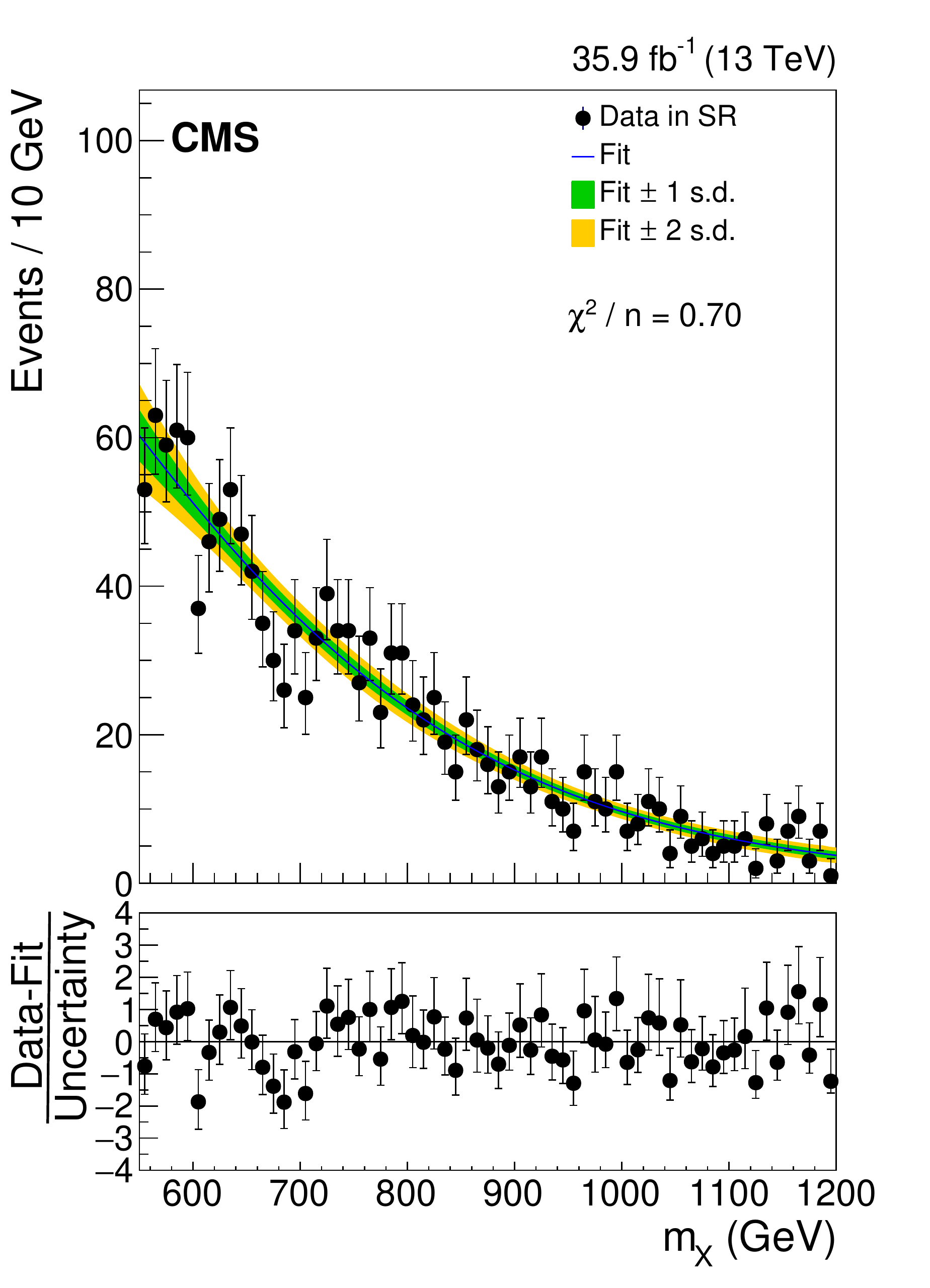}
\caption{The $m_{\mathrm{X}}$ distribution for the multijet background in the SR in data for the MMR. A fit to the background-only hypothesis, which consists of the multijet shape, is shown. The shaded regions correspond to variations of $\pm$1 and $\pm$2 standard deviation (s.d.) in this parametrized form. Here $n$ is the number of degrees of freedom in each fit. The lower panel shows the difference between the data and the fits, divided by the uncertainty in the number of data events.}
\label{fig:MMRfit}
\end{figure}

\begin{figure}[h!]
\centering
\includegraphics[width=0.7\textwidth]{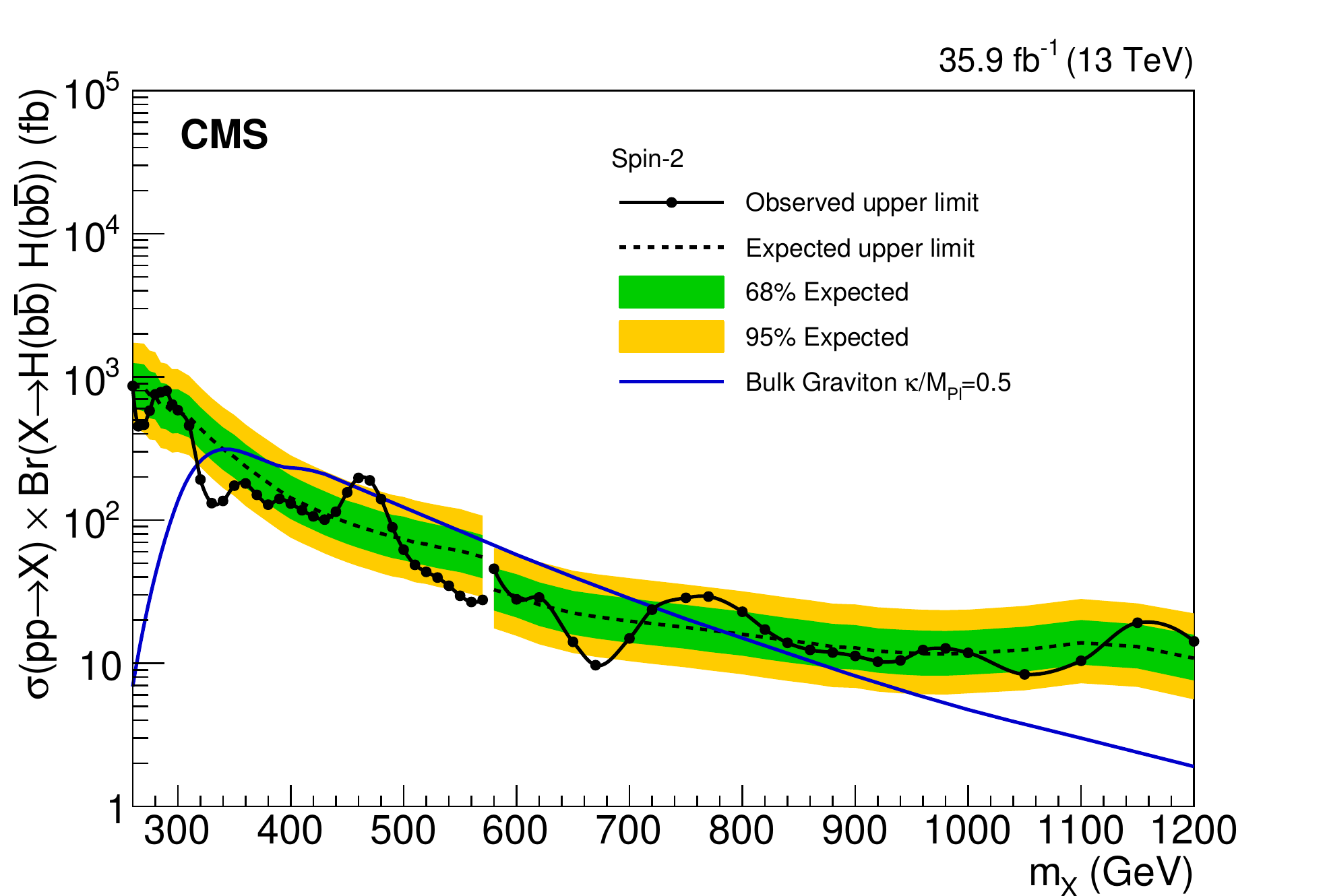}
\caption{The observed and expected upper limits on the cross section for a spin-2 resonance \XHbbHbb at 95\% \CL, using the asymptotic \CLs method. The theoretical cross section for the bulk KK-graviton, with $\kappa/\Mpl= 0.5$ and $\kappa l = 35$, decaying to four \cPqb\ jets via Higgs bosons is overlaid. The transition between the LMR and the MMR is based on the expected sensitivity, resulting in the observed discontinuity. }
\label{fig:UpperLimit_Combined}
\end{figure}

\begin{figure}[h!]
\centering
\includegraphics[width=0.7\textwidth]{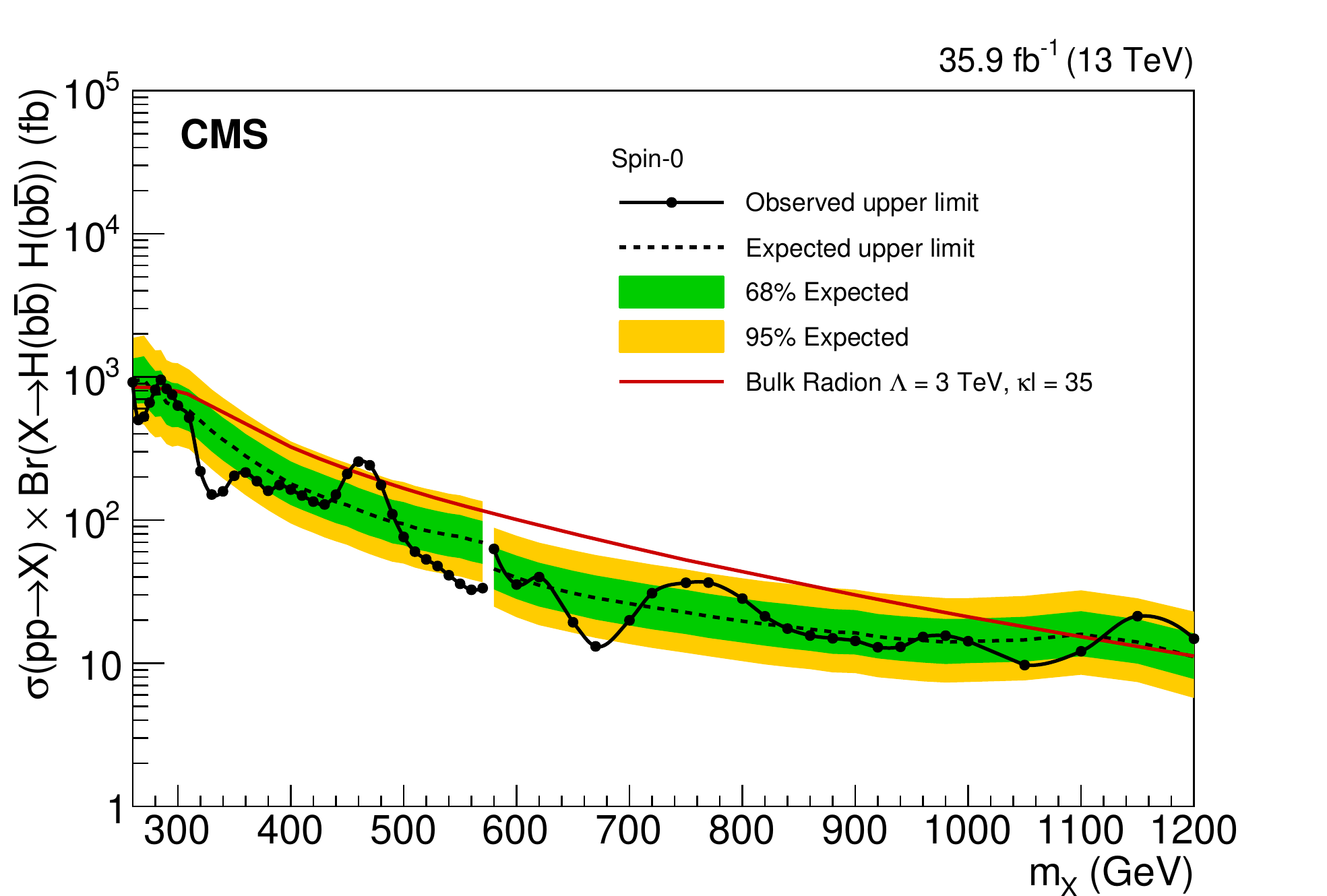}
\caption{The observed and expected upper limits on the cross section for a spin-0 resonance \XHbbHbb at 95\% \CL, using the asymptotic \CLs method. The theoretical cross section for the production of a radion, with $\Lambda=3\TeV$, $\kappa l=35$, and no radion-Higgs boson mixing, decaying to four \cPqb\ jets via Higgs bosons is overlaid. The transition between the LMR and the MMR is based on the expected sensitivity, resulting in the observed discontinuity.}
\label{fig:UpperLimit_Combined_radion}
\end{figure}

The $m_{\mathrm{X}}$ distribution for data in the SR and results of the fit to the parametric background model are shown in Figs.~\ref{fig:unblinded}, for the LMR1 and the LMR2, and \ref{fig:MMRfit} for the MMR. Parameters controlling the shapes and yields of the signal are allowed to float within ranges determined by the systematic uncertainties. The parameters and normalization of the multijet background shape are left to float freely in the fit.
The shapes of the background-only fit are found to adequately describe the data in each of the search region.

The observed and expected upper limits on the cross section times branching fraction for \ppXHbbHbb at 95\% confidence level (\CL) are computed using the modified frequentist \CLs method~\cite{Cowan:2010js,CLS2,CLS1,CMS-NOTE-2011-005}. These limits are shown in Figs.~\ref{fig:UpperLimit_Combined} and \ref{fig:UpperLimit_Combined_radion} for spin-2 and spin-0 hypotheses, respectively. The green and yellow bands represent the $\pm$1 and $\pm$2 standard deviation confidence intervals around the expected limits. The observed limits in these figures exhibit several deviations from the expected one beyond $\pm 2$ standard deviation. The local $p$-value of the most significant excess (deficit) is 2.6 ($-3.6$) standard deviation. In order to estimate the global significance of these deviations within the search range, we estimated, based on pseudo-experiments, a global probability to see three (four) positive or negative deviations from the expected limits consistent with narrow positive or negative signal contributions with the square sum of the significances exceeding 5.1 (5.6) standard deviation, as observed in data. In both cases, the global probability was found to be in a percent range, consistent with a statistical nature of the observed deviations.
In Fig.~\ref{fig:UpperLimit_Combined}, the NLO theoretical cross section for the gluon fusion production of a bulk KK-graviton with $\kappa l$ and $\kappa$ set to 35 and $0.5\Mpl$ respectively is shown. This KK-graviton is excluded at 95\% \CL in the mass ranges of 320--450 and 480--720\GeV.
In Fig.~\ref{fig:UpperLimit_Combined_radion}, the NLO theoretical cross section for the gluon fusion production of a radion with decay constant $\Lambda= 3\TeV$ is shown. Such a radion is excluded at 95\% \CL in the mass ranges of 260--280, 300--450, and 480--1120\GeV.

\section{Summary}
A search for a narrow-width resonance decaying into two Higgs bosons, each decaying into a bottom quark-antiquark pair, is presented. The search is performed using proton-proton collision data corresponding to an integrated luminosity of 35.9\fbinv  at $\sqrt{s} = 13$\TeV recorded by the CMS detector at the LHC. No evidence for a signal is observed and upper limits at 95\% confidence level on the production cross section for spin-0 and spin-2 resonances in the mass range from 260 to 1200\GeV are set. These cross-section limits are translated into an exclusion at 95\% confidence level of a bulk KK-graviton (with $\kappa l$ = 35 and $\kappa = 0.5\Mpl$) in the mass ranges of 320--450\GeV and 480--720\GeV. The corresponding excluded mass ranges for a radion (with decay constant $\Lambda= 3\TeV$) are 260--280\GeV, 300--450\GeV and 480--1120\GeV. This analysis outperforms a similar search by CMS using 17.9\fbinv collected at 8\TeV~\cite{Khachatryan:2015yea} and extends the sensitivity to the gluon fusion production of a radion with decay constant $\Lambda= 3\TeV$ and to bulk graviton with $\kappa$ set to $0.5\Mpl$.

\begin{acknowledgments}
We congratulate our colleagues in the CERN accelerator departments for the excellent performance of the LHC and thank the technical and administrative staffs at CERN and at other CMS institutes for their contributions to the success of the CMS effort. In addition, we gratefully acknowledge the computing centres and personnel of the Worldwide LHC Computing Grid for delivering so effectively the computing infrastructure essential to our analyses. Finally, we acknowledge the enduring support for the construction and operation of the LHC and the CMS detector provided by the following funding agencies: BMWFW and FWF (Austria); FNRS and FWO (Belgium); CNPq, CAPES, FAPERJ, FAPERGS, and FAPESP (Brazil); MES (Bulgaria); CERN; CAS, MoST, and NSFC (China); COLCIENCIAS (Colombia); MSES and CSF (Croatia); RPF (Cyprus); SENESCYT (Ecuador); MoER, ERC IUT, and ERDF (Estonia); Academy of Finland, MEC, and HIP (Finland); CEA and CNRS/IN2P3 (France); BMBF, DFG, and HGF (Germany); GSRT (Greece); NKFIA (Hungary); DAE and DST (India); IPM (Iran); SFI (Ireland); INFN (Italy); MSIP and NRF (Republic of Korea); LAS (Lithuania); MOE and UM (Malaysia); BUAP, CINVESTAV, CONACYT, LNS, SEP, and UASLP-FAI (Mexico); MBIE (New Zealand); PAEC (Pakistan); MSHE and NSC (Poland); FCT (Portugal); JINR (Dubna); MON, RosAtom, RAS, RFBR, and NRC KI (Russia); MESTD (Serbia); SEIDI, CPAN, PCTI, and FEDER (Spain); Swiss Funding Agencies (Switzerland); MST (Taipei); ThEPCenter, IPST, STAR, and NSTDA (Thailand); TUBITAK and TAEK (Turkey); NASU and SFFR (Ukraine); STFC (United Kingdom); DOE and NSF (USA).

\hyphenation{Rachada-pisek} Individuals have received support from the Marie-Curie programme and the European Research Council and Horizon 2020 Grant, contract No. 675440 (European Union); the Leventis Foundation; the A. P. Sloan Foundation; the Alexander von Humboldt Foundation; the Belgian Federal Science Policy Office; the Fonds pour la Formation \`a la Recherche dans l'Industrie et dans l'Agriculture (FRIA-Belgium); the Agentschap voor Innovatie door Wetenschap en Technologie (IWT-Belgium); the F.R.S.-FNRS and FWO (Belgium) under the ``Excellence of Science - EOS" - be.h project n. 30820817; the Ministry of Education, Youth and Sports (MEYS) of the Czech Republic; the Lend\"ulet (``Momentum") Programme and the J\'anos Bolyai Research Scholarship of the Hungarian Academy of Sciences, the New National Excellence Program \'UNKP, the NKFIA research grants 123842, 123959, 124845, 124850 and 125105 (Hungary); the Council of Science and Industrial Research, India; the HOMING PLUS programme of the Foundation for Polish Science, cofinanced from European Union, Regional Development Fund, the Mobility Plus programme of the Ministry of Science and Higher Education, the National Science Center (Poland), contracts Harmonia 2014/14/M/ST2/00428, Opus 2014/13/B/ST2/02543, 2014/15/B/ST2/03998, and 2015/19/B/ST2/02861, Sonata-bis 2012/07/E/ST2/01406; the National Priorities Research Program by Qatar National Research Fund; the Programa Estatal de Fomento de la Investigaci{\'o}n Cient{\'i}fica y T{\'e}cnica de Excelencia Mar\'{\i}a de Maeztu, grant MDM-2015-0509 and the Programa Severo Ochoa del Principado de Asturias; the Thalis and Aristeia programmes cofinanced by EU-ESF and the Greek NSRF; the Rachadapisek Sompot Fund for Postdoctoral Fellowship, Chulalongkorn University and the Chulalongkorn Academic into Its 2nd Century Project Advancement Project (Thailand); the Welch Foundation, contract C-1845; and the Weston Havens Foundation (USA).
\end{acknowledgments}

\bibliography{auto_generated}

\cleardoublepage \appendix\section{The CMS Collaboration \label{app:collab}}\begin{sloppypar}\hyphenpenalty=5000\widowpenalty=500\clubpenalty=5000\input{HIG-17-009-authorlist.tex}\end{sloppypar}
\end{document}

%% file: HIG-17-009-authorlist.tex
\vskip\cmsinstskip
\textbf{Yerevan Physics Institute, Yerevan, Armenia}\\*[0pt]
A.M.~Sirunyan, A.~Tumasyan
\vskip\cmsinstskip
\textbf{Institut f\"{u}r Hochenergiephysik, Wien, Austria}\\*[0pt]
W.~Adam, F.~Ambrogi, E.~Asilar, T.~Bergauer, J.~Brandstetter, E.~Brondolin, M.~Dragicevic, J.~Er\"{o}, A.~Escalante~Del~Valle, M.~Flechl, M.~Friedl, R.~Fr\"{u}hwirth\cmsAuthorMark{1}, V.M.~Ghete, J.~Grossmann, J.~Hrubec, M.~Jeitler\cmsAuthorMark{1}, A.~K\"{o}nig, N.~Krammer, I.~Kr\"{a}tschmer, D.~Liko, T.~Madlener, I.~Mikulec, E.~Pree, N.~Rad, H.~Rohringer, J.~Schieck\cmsAuthorMark{1}, R.~Sch\"{o}fbeck, M.~Spanring, D.~Spitzbart, A.~Taurok, W.~Waltenberger, J.~Wittmann, C.-E.~Wulz\cmsAuthorMark{1}, M.~Zarucki
% ...
\vskip\cmsinstskip
% ...
1:  Also at Vienna University of Technology, Vienna, Austria\\
% ...
